\documentclass{anstrans}

\usepackage{graphicx, booktabs, algorithm, algorithmicx, algpseudocode, microtype}
\usepackage{hyperref, epstopdf}

\algnewcommand\algorithmicestep{\textbf{E-Step:}}
\algnewcommand\EStep{\State\algorithmicestep}
\algnewcommand\algorithmicmstep{\textbf{M-Step:}}
\algnewcommand\MStep{\State\algorithmicmstep}
\newcommand{\ignore}[1]{{}}
\newcommand{\nn}{\nonumber}

\title{Online Diversion Detection in Nuclear Fuel Cycles via Multimodal Observations}
\author{Yasin Y{\i}lmaz$^{1}$, Elizabeth Hou$^{2}$, Alfred O. Hero$^{1}$}

\institute{
$^{1}$University of Michigan, Department of Electrical Engineering and Computer Science, Ann Arbor, MI 48109
\and
$^{2}$University of Michigan, Department of Statistics, Ann Arbor, MI 48109
}

\email{yasiny@umich.edu \and emhou@umich.edu \and hero@eecs.umich.edu}

\begin{document}

\section{Introduction}

In nuclear fuel cycles, an enrichment facility typically provides low enriched uranium (LEU) to a number of customers. We consider monitoring an enrichment facility to timely detect a possible diversion of highly enriched uranium (HEU). To increase the the detection accuracy it is important to efficiently use the available information diversity. In this work, it is assumed that the shipment times and the average power consumption of the enrichment facility are observed for each shipment of enriched uranium.
We propose to initially learn the statistical patterns of the enrichment facility through the bimodal observations in a training period, that is known to be free of diversions. Then, for the goal of timely diversion detection, we propose to use an online detection algorithm which sequentially compares each set of new observations in the test period, which possibly includes diversions, to the learned patterns, and raises a diversion alarm when a significant statistical deviation is detected. The efficacy of the proposed method is shown by comparing its detection performance to those of the traditional detection methods in the Statistics literature. 

In the remainder of the paper, firstly, the assumed nuclear fuel cycle model is given; secondly, the traditional detection algorithms and the proposed algorithm are presented; thirdly, the simulation results are provided; and finally, the paper is concluded.

\section{Observation Model}

\begin{figure}[thb] 
  \centering
  \includegraphics[width=\linewidth]{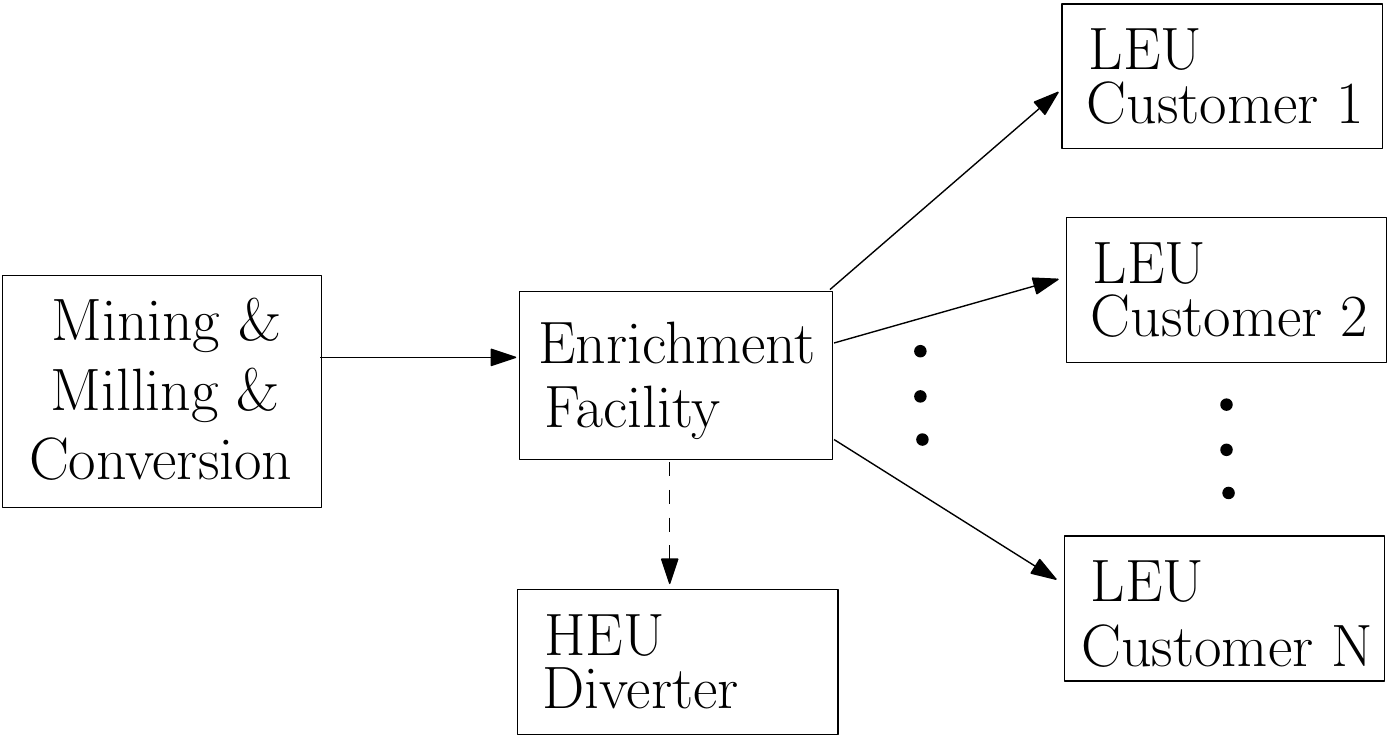}
  \caption{Considered nuclear fuel cycle.}
  \label{fig:NFC}
\end{figure}

The considered  nuclear fuel cycle diagram is shown in Figure \ref{fig:NFC}. An enrichment facility, which serves to an unknown number of LEU customers, and possibly diverts uranium to an HEU diverter, is monitored. For each shipment from the enrichment facility, we observe the time elapsed since the previous shipment, which is used as a proxy for the duration of production of shipped materials, as well as the average power consumption for this period. We assume that the observed average power consumption of the facility is representative for the average power consumption used for enrichment, which is measured by MTSWU/day. Sample observations, generated from the IAEA Nuclear Fuel Cycle Simulation System (NFCSS) \cite{IAEA}, are given in Table \ref{tab:ship}. 

\begin{table}[thb]
\caption{Sample shipments of $1$ ton with duration and average power consumption observations.}
\label{tab:ship}
\begin{tabular}{| p{0.3\linewidth} | p{0.2\linewidth} | p{0.3\linewidth} |}
\hline
{\bf Shipment no. \newline \includegraphics[scale=0.13]{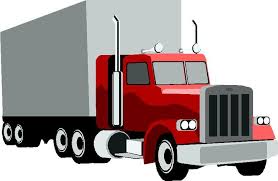}} & {\bf Duration \newline (days)} & {\bf Average Power \newline Consumption \newline (MTSWU/day)}  \\ 
\hline
\centering 1 & 17.11 & 0.2015 \\
\hline
\centering 2 & 43.33  & 0.1018 \\
\hline
\end{tabular}
\end{table}

The shipments (e.g., trucks) that leave the facility can be observed by either a human or a device such as camera or satellite. The average power consumption may be measured by means of electricity meters or estimated from other observations such as smoke activity. 

Each customer may request a different enrichment level between 3\% and 5\%, and a different delivery mode (e.g., standard or expedited), which complicates the statistical patterns of the enrichment facility. The delivery mode determines the average power consumption of the facility; and the requested enrichment level determines the total energy the facility needs to consume to produce the requested amount of uranium enriched at that level. For instance, according to IAEA NFCSS, $3.43$ MTSWU energy is needed to produce $1$ ton of $3\%$ enriched uranium. 
If the facility runs at the average power levels of $0.1$ MTSWU/day and $0.2$ MTSWU/day at the standard and expedited delivery modes, respectively, then it would take $34.3$ days at the standard mode, and $17.15$ days at the expedited mode for the facility to produce $1$ ton of $3\%$ enriched uranium.

\begin{figure}[thb] 
  \centering
  \includegraphics[width=\linewidth]{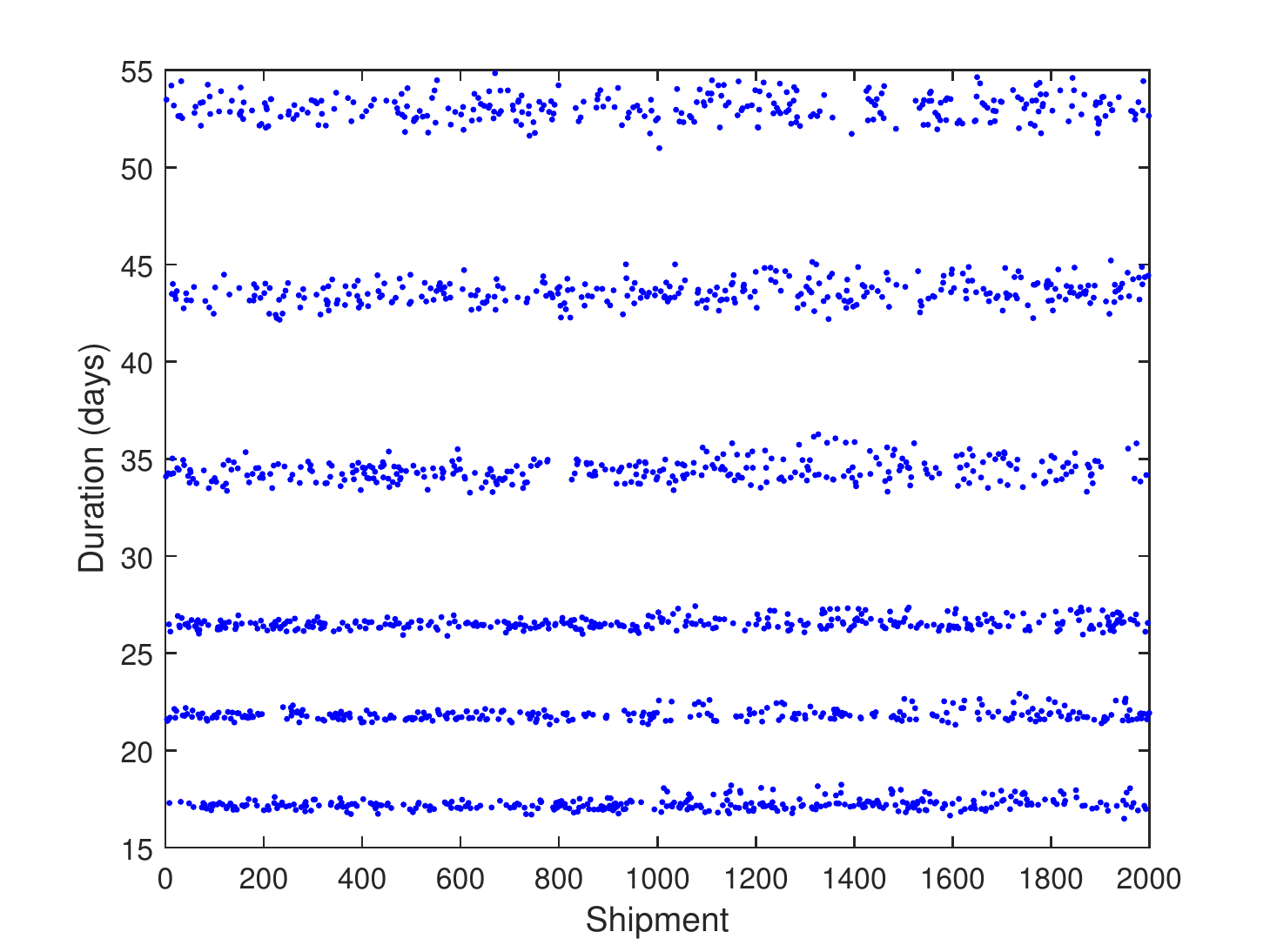}
  \includegraphics[width=\linewidth]{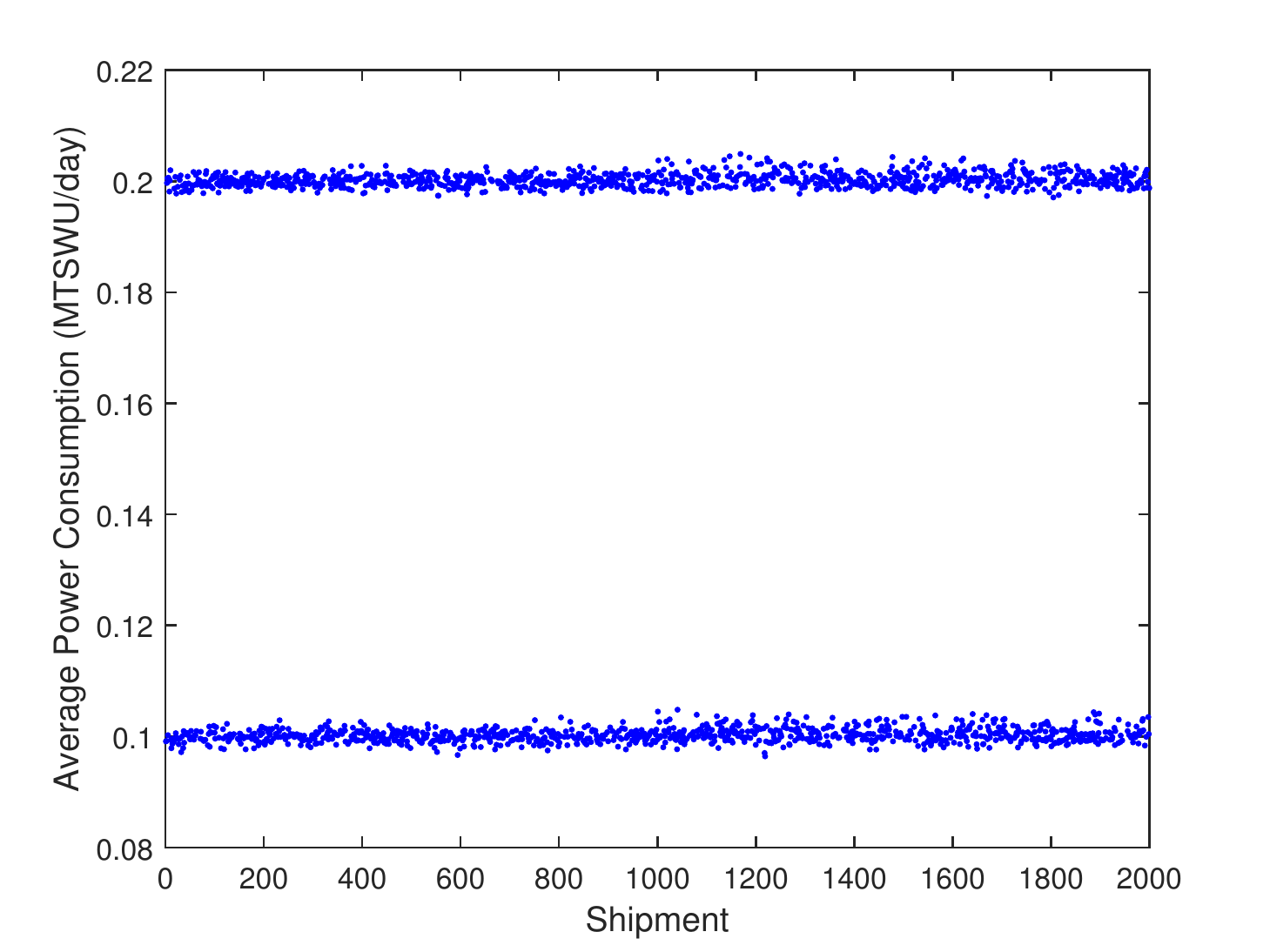}
  \caption{Sample observations from a scenario in which there are $6$ customer patterns (as a result of $3$ different enrichment levels and $2$ different delivery modes), and $1$ kg of $90\%$ HEU is diverted on average once every $5$ shipments after the $1000$th shipment. The enrichment facility increases its average power consumption by $1\%$ on average during the shipments that include diversion; that way it avoids a significant increase in the shipment durations, and as a result the detection-by-visual-inspection of its illicit activity.}
  \label{fig:obs}
\end{figure}

In the diversion scenarios considered in this paper, the facility illicitly produces a small amount of HEU (e.g., $1$ kg of $90\%$ enriched uranium), in addition to its regular LEU production. Hence, when a diversion occurs, we assume the total energy consumption of the facility will slightly increase, which will be manifested in our observations as a small increase in shipment duration and/or average power consumption. Figure \ref{fig:obs} shows sample duration and average power consumption observations for the diversion scenario in which both observations slightly increase after the thousandth shipment. It is seen that such small changes cannot be detected by visual inspection.
Using statistical methods we aim to detect such small changes in a timely manner. 

\section{Proposed Detection Algorithms}

In multivariate problems, assuming independence among the variates (i.e., data dimensions), it is common to apply univariate analysis separately in each dimension, mainly for simplicity. However, when the independence assumption does not hold, separate univariate analysis inevitably suffers significant performance loss. For the problem under consideration, we will show that this is actually the case, and we need more sophisticated \emph{joint} analysis to obtain acceptable detection performances. We will present four statistical detection tests in this section, and compare their performances through simulations in the next section.

\subsection{Kolmogorov-Smirnov Test}

We start with the Kolmogorov-Smirnov (KS) test \cite[page 95]{Kolmog}, a nonparametric detection technique which in the training phase estimates the cumulative distribution function (cdf) for the duration and power observations separately to use as a baseline in the test phase. We use the Gaussian kernel function with the optimum bandwidth for cdf estimation. Let $F(x)$ and $G(x)$ denote the estimated cdf of the duration and power observations, respectively. In the test phase, for a window of observations, similarly cdf is estimated via Gaussian kernel and compared to the baseline cdf to detect a possible significant deviation. We continue the test phase by sliding the window in time until a significant change in cdf is detected. Using a window of size $W$ we denote the estimated cdf with $F^W_n(x)$ and $G^W_n(x)$ in the $n$-th window. The test statistics and the stopping time are given by
\begin{align}
	D_n &= \sup_x |F^W_n(x)-F(x)|, ~~ 	E_n = \sup_x |G^W_n(x)-G(x)| \nn\\
	T^K_d &= \min\{n: D_n\ge \delta\}, ~~ T^K_p = \min\{n: E_n\ge \delta\} \nn\\
	T^K &= \min\{T^K_d,T^K_p\},
\end{align}
where $\sup$ denotes the supremum operator in mathematics.

At each time frame $n$, the test statistics $D_n$ and $E_n$ are computed for the duration and power observations, respectively; and compared to the threshold $\delta$. The first time either $D_n$ or $E_n$ exceeds $\delta$, the algorithm stops and raises a diversion alarm. The threshold $\delta$ controls a trade-off between the false alarm rate and the detection delay. In particular large $\delta$ ensures small false alarm rate, but at the same time causes large detection delays. Similarly, the window size $W$ controls a trade-off between the false alarm rate and the detection delay: small $W$ yields smaller detection delays at the expense of less accurate cdf estimation, and thus increased false alarm rate. 

The KS test is preferred over the chi-square test for its ability to work with smaller window sizes (i.e., higher time resolution) \cite{Kolmog}. Nevertheless, since $W\gg1$, it is not a completely online detection technique, i.e., it does not make a detection decision after each new observation arrives.

\subsection{CUSUM Test}

As opposed to the KS test, the cumulative sum (CUSUM) test is a truly online method which recursively updates its test statistic at each time $t$ as follows
\begin{equation}
\label{eq:CUSUM}
	S_t = \max\left\{S_{t-1}+\log\frac{f_1(x_t)}{f_0(x_t)}\right\}, ~~ S_0 = 0, 
\end{equation}
where $f_0(x_t)$ and $f_1(x_t)$ are the pre-change and post-change probability density function (pdf) of the data sample $x_t$. Similar to KS, CUSUM stops the first time $S_t$ exceeds a threshold, i.e., at the stopping time
\begin{equation}
	T^C = \min\{t: S_t \ge \rho\}. \nn
\end{equation}
Similar to $\delta$ in KS, the threshold $\rho$ is selected to strike a balance between the false alarm rate and the detection delay. 

As opposed to KS, CUSUM does not forget its change history, but accumulates it through time, which makes CUSUM a very powerful online detection technique. In fact, when both $f_0(x)$ and $f_1(x)$ are completely known, CUSUM is optimum in terms of minimizing the expected detection delay in the worst-case scenario (i.e., in the minimax sense) \cite{CUSUM}. The challenge in using CUSUM lies in estimating the post-change pdf $f_1(x)$. Due to online testing, typically a parametric model is assumed and its parameters are estimated at each time following the maximum likelihood approach. This framework is known as the generalized CUSUM, and it is asymptotically optimum in terms of minimizing the expected detection delay in the worst-case scenario \cite{CUSUM}. 

The most straightforward way to apply generalized CUSUM to our diversion detection problem is to fit Gaussian pdfs to the pre-change and post-change duration and power data. Denote the pre-change and post-change Gaussian pdfs with $f^G_{0d}$ and $f^G_{1d}$ for the duration data, and with $f^G_{0p}$ and $f^G_{1p}$ for the power data, respectively. Two parallel CUSUMs are run for the duration and power datasets with the following test statistics
\begin{align}
\begin{split}
\label{eq:CUSUM_stat}
	U^G_t &= \max\left\{U^G_{t-1}+\log\frac{f^G_{1d}(x_t)}{f^G_{0d}(x_t)}\right\}, ~~ U^G_0 = 0, \\
	V^G_t &= \max\left\{V^G_{t-1}+\log\frac{f^G_{1p}(x_t)}{f^G_{0p}(x_t)}\right\}, ~~ V^G_0 = 0, 
\end{split}
\end{align}
and the stopping times
\begin{align}
\label{eq:G-CUSUM}
	T^G_d &= \min\{t: U^G_t \ge \rho^G\}, \nn\\
	T^G_p &= \min\{t: V^G_t \ge \rho^G\}, \nn\\
	T^G &= \min\{T^G_d,T^G_p\},
\end{align}
where $U^G_t$ and $V^G_t$ are given by \eqref{eq:CUSUM_stat}. We call the Gaussian-based CUSUM given in \eqref{eq:G-CUSUM} as G-CUSUM. At each time $t$, the test statistics $U^G_t$ and $V^G_t$ are computed for the duration and power observations, respectively; and compared to the threshold $\rho^G$. The first time either $U^G_t$ or $V^G_t$ exceeds $\rho^G$, the algorithm stops and raises a diversion alarm. The threshold $\rho^G$ controls a trade-off between the false alarm rate and the detection delay.

From the problem definition, we know that each customer has its own characteristic enrichment level and duration demands. Hence, the observed datasets are actually generated from a mixture of a number of probability distributions. Assuming that only one customer is served in each shipment we can fit Gaussian mixture models to the duration and power datasets separately. Similar to the derivation of G-CUSUM, given in \eqref{eq:G-CUSUM}, we can write the test statistics and the stopping time for the Gaussian mixture model-based CUSUM (GM-CUSUM). 

\begin{figure}[t] 
  \centering
  \includegraphics[width=\linewidth]{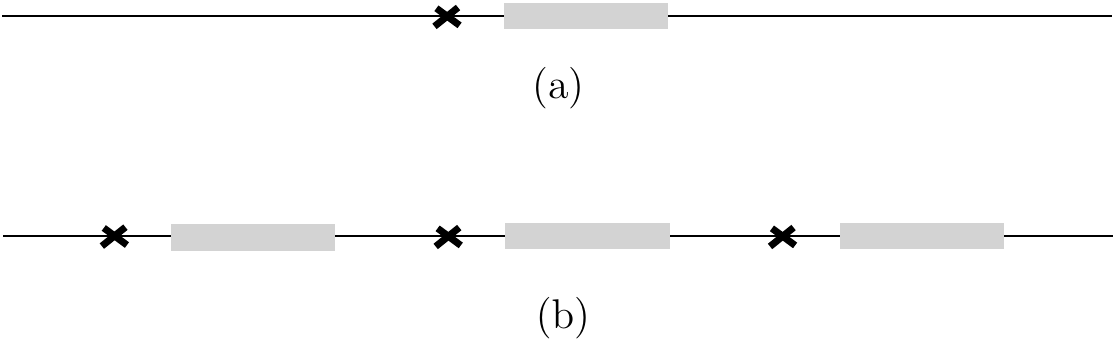}
  \caption{The mean-shift model to derive $f_1(x)$ from $f_0(x)$ for (a) G-CUSUM and (b) GM-CUSUM. Estimated means for $f_0(x)$ are marked with x. In (b), feasible set, denoted with $\mathcal{A}$ in \eqref{eq:mean}, consists of the points marked with x for $f_0(x)$ and the shaded regions for $f_1(x)$. Similarly in (a).}
  \label{fig:mean}
\end{figure}

Both in G-CUSUM and GM-CUSUM, the parameters (mean and variance) of Gaussians that fit $f_0(x)$ are estimated from the training data, and for $f_1(x)$ a small to moderate shift is assumed in the mean with respect to $f_0(x)$, as shown in Figure \ref{fig:mean}. No change is assumed in the variance. While computing the test statistics in test phase, the mean is selected from the feasible set according to the maximum likelihood criterion, i.e., 
\begin{equation}
\label{eq:mean}
	\mu^* = \arg\max_{\mu \in \mathcal{A}} f^G(x|\mu,\sigma),
\end{equation}
where $f^G$ is the Gaussian pdf, and $\mathcal{A}$ denotes the feasible set for $\mu$.

\subsection{Multimodal CUSUM Test}

We can improve the performance of GM-CUSUM by jointly modeling the duration and power data. Denoting the duration and power data with $y_t$ and $z_t$, respectively, and the number of distinct customer patterns with $M$ we write, for the $m$-th customer pattern, the joint pdf as
\begin{equation}
\label{eq:joint}
	f_m(y_t,z_t) = f_m(y_t|z_t) f_m(z_t), ~~ m=1,\ldots,M.
\end{equation}
Note that the duration is given by the relation $y_t=e_t/z_t$ where $e_t$ is the consumed energy. Hence, given the power $z_t$, the uncertainty in $y_t$ is due to only $e_t$, and accordingly we can write $f_m(y_t|z_t)=f_m(e_t)$, where $e_t=y_t z_t$. Then, \eqref{eq:joint} becomes
\begin{equation}
	f_m(y_t,z_t) = f_m(e_t) f_m(z_t), ~~ m=1,\ldots,M.
\end{equation}
In this case, from \ref{eq:G-CUSUM} and the generalized CUSUM approach, the test statistic and the stopping time for the multimodal CUSUM (M-CUSUM) are given by
\begin{align}
\label{eq:M_stat}
	S^M_t &= \max\left\{S^M_{t-1}+\log\frac{\max_{m=1,\ldots,M}  f_{1m}(e_t) f_{1m}(z_t)}{\max_{m=1,\ldots,M}  f_{0m}(e_t) f_{0m}(z_t)}\right\}, ~~ S^M_0 = 0. \\
\label{eq:M_stop}
	T^M &= \min\{t: S^M_t \ge \rho^M\}.
\end{align}

Now the challenge is how to estimate $f_{0m}(e_t)$ and $f_{0m}(z_t)$. We can use the mean-shift model described in \eqref{eq:mean} and Figure \ref{fig:mean} to model $f_{1m}(e_t)$ and $f_{1m}(z_t)$ from $f_{0m}(e_t)$ and $f_{0m}(z_t)$, respectively. To estimate $f_{0m}(e_t)$ and $f_{0m}(z_t)$ we propose to first cluster the shipments, $t=1,\ldots,\tau$, where $\tau$ is the length of the training dataset. For clustering, the straightforward way is to apply the k-means algorithm to the data matrix $[y_t, z_t]_{t=1,\ldots,\tau}$ directly. However, the range and the scale of $x_t$ and $y_t$ might be very different, as seen in Table \ref{tab:ship}. Using the multimodal factor analysis (MMFA) \cite{MMFA} method  we first obtain a unified representation $w_t$ for each shipment $t$. Then, we cluster the shipments using k-means on $w_t$, and fit a Gaussian, in each cluster $m$, for $e_t$ and $p_t$ to estimate $f_{0m}(e_t)$ and $f_{0m}(z_t)$, respectively. The number of clusters $M$ is selected by trying some numbers and selecting the one with the highest silhouette value. The M-CUSUM procedure is summarized in Algorithm \ref{alg:M-CUSUM}.

\begin{algorithm}[htb]
\caption{The proposed M-CUSUM procedure}
\label{alg:M-CUSUM}
\baselineskip=0.5cm
  \begin{algorithmic}[1]
  \State {\bf Training data}
    \State ~~~Obtain $[w_t]$ from $[y_t, z_t]$ using MMFA
    \State ~~~Cluster $[w_t]$ using k-means and silhouette values
    \State ~~~Fit Gaussians to each cluster to estimate $f_{0m}(e_t)$, $f_{0m}(z_t)$
    \State ~~~Estimate $f_{1m}(e_t)$, $f_{1m}(z_t)$ as in  Figure \ref{fig:mean}
  \State {\bf Test data}
    \State ~~~Compute test statistic as in \eqref{eq:M_stat} and stop according to \eqref{eq:M_stop}
  \end{algorithmic}
\end{algorithm}

\section{Simulation Results}

\begin{figure}[thb] 
  \centering
  \includegraphics[width=1.1\linewidth]{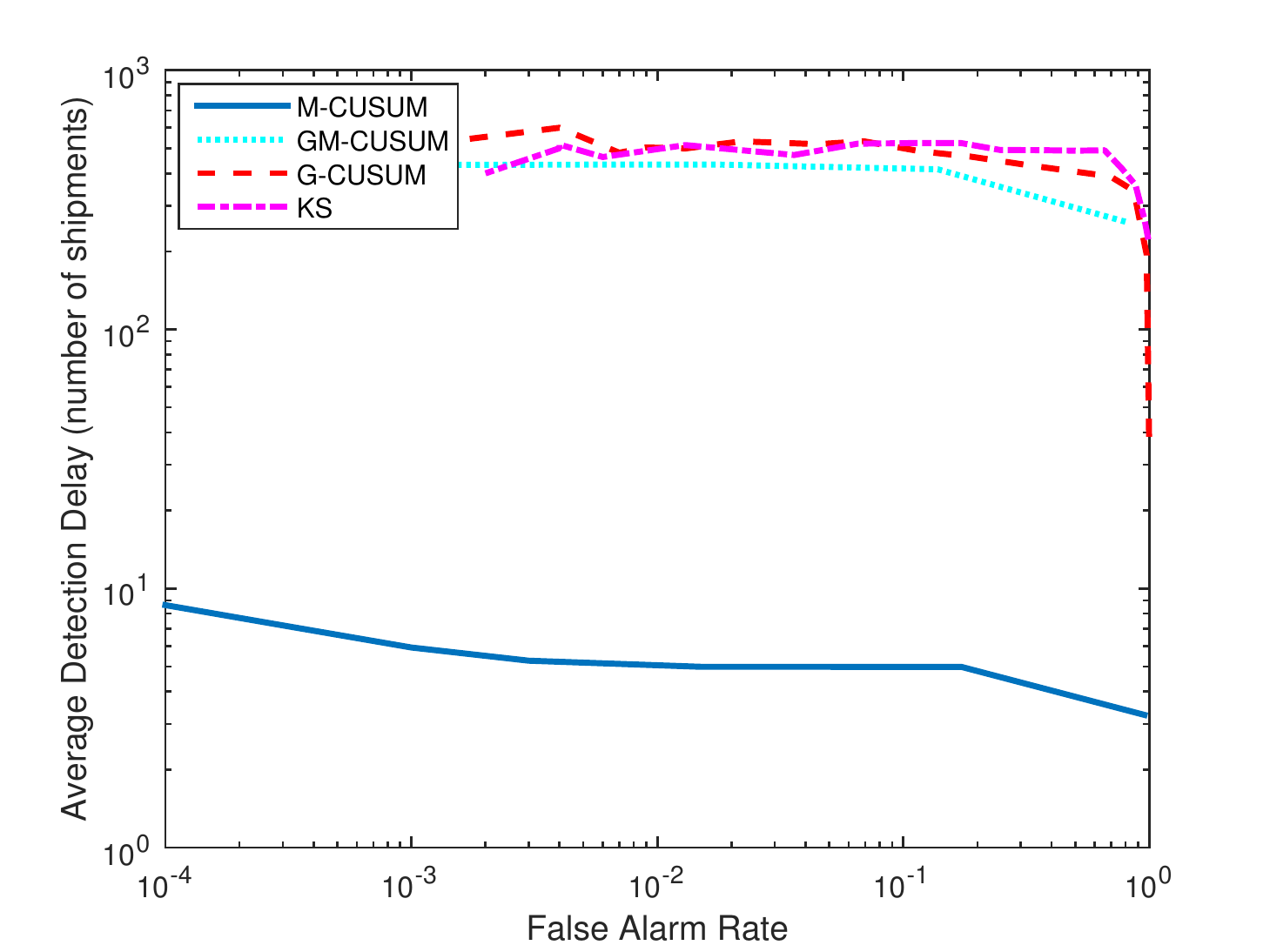}
  \caption{The proposed M-CUSUM algorithm, which jointly analyzes the bimodal data shown in Figure \ref{fig:obs}, significantly outperforms the conventional detection algorithms which separately analyze each data modality. Over the entire range of false alarm rates, M-CUSUM significantly reduces the average delay to detect diversions. Results are obtained by averaging over 1000 trials.}
  \label{fig:add}
\end{figure}

In this section, we provide simulation results to compare the four detection algorithms discussed in the previous section. 
In our simulations, we generated data from $6$ customer patterns: $2$ in power levels with means $0.1$ MTSWU/day and $0.2$ MTSWU/day and standard deviation $0.001$ MTSWU/day, and $3$ in enrichment levels ($3\%$, $3.5\%$ and $4\%$).
From the IAEA NFCSS website,  we have set the means for energy levels (MTSWU) as $3.43$ for $3\%$, $4.35$ for $3.5\%$, and $5.29$ for $4\%$, and the standard deviation as $0.03$. A customer identity is generated for each shipment in the training and test datasets from multinomial distributions with uniform probabilities for both energy and power. Next, for each shipment, energy and power levels are generated from the Gaussian distributions that correspond to the assigned customer pattern (e.g., means $3.43$ MTSWU and $0.1$ MTSWU/day and standard deviations $0.001$ and $0.03$ for energy and power, respectively). 
Training data length was 1000. In the test dataset, diversions occur with a probability of $0.2$. At each diversion, $1$ kg $90\%$ enriched uranium is produced in addition to the regular LEU production. From the IAEA NFCSS website, this corresponds to an increase of $0.1934$ MTSWU in the energy level. We consider the most challenging diversion scenario: the enrichment facility increases its average power consumption by one standard deviation on average (i.e., means $0.101$ and $0.201$ MTSWU/day) during the shipments that include diversion; that way it avoids a significant increase in the shipment durations.

Figure \ref{fig:add} shows the average detection delay, $E[T-T^*|T>T^*]$, where $E[\cdot]$ is the expectation, $T^*$ is the first shipment diversion occurs, and $T$ is the stopping shipment of the algorithm, against the false alarm rate, $P(T<T^*)$, where $P(\cdot)$ is the probability. As expected M-CUSUM, the proposed algorithm which jointly processes the multimodal observations, significantly outperform the others. Actually, M-CUSUM is the only algorithm that detects the diversion with a reasonable delay.
GM-CUSUM performs better than KS and G-CUSUM because it fits a finer model to the data. G-CUSUM performs poorly because its unimodal Gaussian model fails to fit well to the complicated statistical pattern of the data. KS has the inherent drawback of not being an online detection technique.

\section{Conclusion}

We have considered the online detection of HEU diversions in nuclear fuel cycles. The proposed detection technique which jointly processes the collected bimodal data was shown to greatly outperform the conventional detection techniques which process each data modality separately. Under the challenging diversion scenario in which there is only a small increase in each data modality, the proposed algorithm is indeed the only algorithm that can detect the diversion with a reasonable delay.

\section{Acknowledgments}
\noindent This work was funded by the Consortium for Verification Technology under Department of Energy National Nuclear Security Administration award number DE-NA0002534.
The authors would like to thank Dr. Meghan McGarry and Dr. Paul Wilson from the Computational Nuclear Engineering Group at the University of Wisconsin-Madison for the fruitful discussions and comments.



\begin{thebibliography}{}
%
%
\bibitem{IAEA}
INTERNATIONAL ATOMIC ENERGY AGENCY,  ``Nuclear Fuel Cycle Simulation System", \url{https://infcis.iaea.org/NFCSS/NFCSSMain.asp?RightP=Calculation&EPage=2}

\bibitem{Kolmog}
R.M. FELDMAN and C. VALDEZ-FLORES,  {\it Applied Probability and Stochastic Processes}, Springer (2009)

\bibitem{CUSUM}
M. BASSEVILLE and I.V. NIKIFOROV,  {\it Detection of Abrupt Changes: Theory and Application}, Prentice Hall (1993)

\bibitem{MMFA}
Y. YILMAZ and A.O. HERO,  ``Multimodal Factor Analysis", {\it IEEE International Workshop on Machine Learning for Signal Processing}, (2015)

\end{thebibliography}

\end{document}